\documentclass[aps,twocolumn,showpacs]{revtex4}
\usepackage{epsfig}
\usepackage{graphicx}
\usepackage{amsmath}
\usepackage{mathrsfs}
\usepackage{dcolumn}
\newcommand{\beq}{\begin{eqnarray}}
\newcommand{\eeq}{\end{eqnarray}}
\newcommand{\be}{\begin{equation}}
\newcommand{\ee}{\end{equation}}
\newcommand{\bey}{\begin{eqnarray}}
\newcommand{\eey}{\end{eqnarray}}
\newcommand{\ba}{\begin{array}}
\newcommand{\ea}{\end{array}}
\newcommand{\bi}{\begin{itemize}}
\newcommand{\ei}{\end{itemize}}
\newcommand{\bem}{\begin{enumerate}}
\newcommand{\eem}{\end{enumerate}}
\newcommand{\bw}{\begin{widetext}}
\newcommand{\ew}{\end{widetext}}
\newcommand{\ra}{\rangle}
\newcommand{\la}{\langle}
\newcommand{\pp}{\partial}
\newcommand{\ov}{\overline}

\newcommand{\ww}{\widetilde}

\newcommand{\bp}{{\bf p}}

\newcommand{\br}{{\bf r}}

\begin{document}

 \title{
 Semiclassical Approach to Survival Probability at Quantum Phase Transitions
  }

 \author{ Wen-ge Wang$^*$, Pinquan Qin, Lewei He, and Ping Wang}

 \affiliation{
 Department of Modern Physics, University of Science and Technology of China,
 Hefei, 230026, China
}

 \date{\today}

 \begin{abstract}

 We study the decay of survival probability at quantum phase transitions (QPT)
 with infinitely-degenerate ground levels at critical points.
 For relatively long times,
 the semiclassical theory predicts power law decay of the survival probability
 in systems with $d=1$ and exponential decay in systems with sufficiently
 large $d$, where $d$ is the degrees of freedom of the classical counterpart of the system.
 The predictions are checked numerically in four models.

 \end{abstract}

 \maketitle


  \section{Introduction}

\begin{figure}
  \vspace{-1.cm}
 \includegraphics[width=7.0cm]{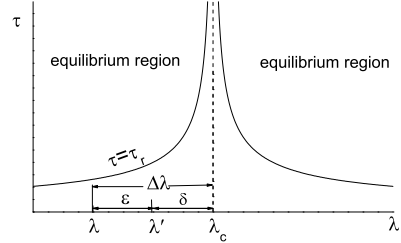}
  \vspace{-0.7cm}
 \caption{
 A schematic plot, where $\tau$ is the time scale of interest
 and $\tau_r$ is the relaxation time, $\lambda $ is a controlling parameter
 with critical value $\lambda_c$ of a QPT.
 Below the solid curves, $\tau <\tau_r$, the system is usually not in an
 equilibrium state and its unitary dynamics should be considered.
 }
 \label{fig-sche}
 \end{figure}

 A quantum phase transition (QPT) is characterized by non-analyticity of the ground
 level of the system at the critical point in the large size limit.
 At a QPT, certain fundamental properties of the ground state (GS) change drastically
 under small variation of a controlling parameter, e.g., strength of a magnetic field.
 Most of the works in QPT have focused on
 properties of equilibrium states (including GS at zero temperature) \cite{Sach99}.
 While, the non-analyticity influences in fact both equilibrium and non-equilibrium
 properties.
 Indeed, when the time scale of interest is smaller than the relaxation time,
 which diverges at the critical point,
 usually the system is not in an equilibrium state and unitary dynamics should be considered
 (Fig.\ref{fig-sche}).
 Due to significant progress in cold atom experiments, time dependent simulation
 of models undergoing QPT is becoming realizable  \cite{Sadler06,Lewen07},
 hence, investigation in the unitary dynamics at QPT is of interest
 both theoretically and experimentally.
 For example,
 resorting to theoretical technique such as a quantum version of the Kibble-Zurek theory
 \cite{Kibble,Zurek85}, it has been shown that
 slow change of the controlling parameter
 passing the critical point may induce some intriguing effects \cite{dyn-qpt}.

 In this paper, we study a different dynamics at QPT,
 which is induced by a sudden small change in the controlling
 parameter, $\lambda \to \lambda'$, in the vicinity of a critical point $\lambda_c$.
 A measure of the effect of this dynamics is the survival probability (SP) of
 an initial state prepared in the GS $|0_\lambda\ra $ of $H(\lambda)$,
 \be M(t) =|\la 0_\lambda |e^{-iH(\lambda')t/\hbar}|0_\lambda\ra |^2. \ee
 The SP, sometimes called autocorrelation function, is a quantity accessible experimentally
 \cite{experiment-SP}.
 Recent it was found that relatively significant and fast decay of the SP
 may indicate the position of QPT \cite{Quan06,LE-qpt,Rossini07},
 which has been demonstrated experimentally \cite{Peng08}.
 Short time decay of the SP has been studied in these works.
 Of further interest, while still unknown, is the law for relatively-long-time
 decay of the SP at QPT and whether it
 may be useful in revealing characteristic properties of QPT \cite{foot-osc,foot-deco}.

 To find an answer to the above question, here we focus on those QPT,
 at the critical points of which the ground levels
 have infinite degeneracy in the large size limit.
 This is a type of QPT met in many cases (see models discussed below and those in
 Ref.~\cite{Sach99}).
 At such a QPT, the non-analyticity may be a consequence of avoided crossings of
 infinite levels, not a few levels.

 We find that the semiclassical theory may be used in the study of the SP decay
 when $\lambda'$ is sufficiently close to $\lambda_c$.
 The theory predicts a power law decay of the SP
 in some systems and an exponential decay in some other systems.
 Numerical results obtained in four models confirm these predictions.



 \section{Semiclassical approach}

 We first discuss a condition for the applicability of the semiclassical theory in the study of the
 SP of GS.
 We use notations:
 $\epsilon =\lambda'-\lambda, \delta =\lambda'-\lambda_c,  \Delta \lambda = \lambda-\lambda_c $
 (Fig.\ref{fig-sche}), and $ \eta = \epsilon / \Delta \lambda$.
 We use $|\alpha_\lambda\ra $ with $\alpha =0,1,\ldots$ to denote eigenstates of
 $H(\lambda)$ with eigenenergies $E_{\alpha}(\lambda )$ in increasing energy order.
 When the ground level of $H(\lambda_c)$ is infinitely degenerate and those of
 $H(\lambda')$ are non-degenerate (or have finite degeneracy),
 infinitely many low-lying levels of $H(\lambda')$ must join its ground level
 in the limit $\lambda' \to \lambda_c$, i.e.,
 \be\lim_{\lambda' \to \lambda_c} E_\alpha(\lambda') = E_0(\lambda_c),
 \hspace{1cm} \text{for many} \ \alpha . \label{E-de} \ee
 This has two consequences:
 (i) $H(\lambda')$ of $\lambda'$ sufficiently close to $\lambda_c$ must have a high density of states
 near its ground level.
 (ii) For a fixed $\lambda $ near $\lambda_c$, when $\lambda'$ is sufficiently close to $\lambda_c$,
 $H(\lambda')$ may have many levels below $\ov E$,
 where $\ov E=\la 0_\lambda|H(\lambda')|0_\lambda\ra $,
 i.e., the initial state $|0_\lambda \ra $ may have a relatively high mean energy
 in the system $H(\lambda')$.
 This is in agreement with
 a property revealed in recent study of the fidelity of GS near critical points,
 which has close relationship to the SP,
 namely, for a fixed small $\epsilon$, the overlap $|\la 0_\lambda|0_{\lambda'}\ra |$
 decreases significantly when $\lambda'$ approaches $\lambda_c$ \cite{GS-fid}.


 Moreover, suppose the system has a classical counterpart in the low energy region.
 Here, a classical counterpart means a classical system, the quantization
 of which gives a system mathematically equivalent to the original quantum system;
 its components are not required to be directly related to components of the original system.
 The property (\ref{E-de}) implies that in the process $\lambda' \to \lambda_c$ longer and longer
 trajectories in the classical system may be of relevance.
 For a fixed initial state $|0_\lambda\ra$, one may assume that the initial value of
 the Lagrangian $L$ does not change notably in this process.
 Then, trajectories of relevance may have large action $S=\int_0^t L dt'$
 for $\lambda'$ sufficiently close to $\lambda_c$.

 The above discussed properties for $\lambda'$ sufficiently close to $\lambda_c$,
 namely, high density of states, relative highness of $\ov E$,
 and large action of some relevant classical trajectories,
 imply that a semiclassical approach may be valid.
 To be specific, for any given $\lambda$ near $\lambda_c$, it is reasonable to expect that
 the semiclassical theory may be used in the study of the SP
 when $\lambda'$ is sufficiently close to $\lambda_c$.


 According to the semiclassical theory, qualitative difference in classical
 trajectories may have quantum manifestation.
 Specifically, in the case of $d=1$
 where $d$ is the degree(s) of freedom of the classical counterpart in the configuration space,
 the classical motion may show periodicity within a time scale
 of interest; on the other hand, for a large $d$, even in a regular system,
 classical trajectories may show no signature of periodicity within times of practical interest.
 This difference suggests that the SP decay in the former case may be
 slower than in the latter case, which we discuss below.

 We consider small $\epsilon $, such that $H(\lambda') = H(\lambda)+\epsilon V$,
 with $V \simeq \frac{dH(\lambda)}{d\lambda}$.
 The SP of the GS of $H(\lambda)$ is a special case of the so-called quantum Loschmidt echo
 or (Peres) fidelity \cite{Peres84}, $M_L(t) = |m(t)|^2 $, where
 \be m(t) = \la \Psi_0|{\rm exp}(iH(\lambda')t/ \hbar ) {\rm exp}(-iH(\lambda)t / \hbar)
 |\Psi_0 \ra . \label{mat} \ee
 In studying the SP, one may employ a semiclassical approach that
 has been found successful in the study of Loschmidt echo
 \cite{JP01,PZ02,CT02,VH03,wwg-LEc,WL05,wwg-LEr,JAB03}.
 For an initial Gaussian wave packet, narrow in
 the coordinate space with width $\sigma$ and centered at ($\ww \br _0,\ww {\bf p}_0$)
 in the phase space,
 using the semiclassical Van Vleck-Gutzwiller propagator, it has been shown that \cite{JP01,VH03}
 \bey m_{\rm sc}(t)  \simeq \left (\pi w^2  \right )^{-d/2}
 \int  d{\bp_0}  \exp { \left [ \frac{i}{\hbar} \Delta S
 - \frac{(\bp_0 - \ww \bp_0 )^2}{w^2} \right ] }
 \label{mt-sc} \eey
 for small $\epsilon $,
 which works in both regular and chaotic cases \cite{VH03,wwg-LEr}.
 Here, $\Delta S$ is the action difference between two nearby trajectories
 in the two systems starting at $(\bp_0 , \ww \br_0 )$ and
 approximately can be evaluated along one trajectory,
 $\Delta S\simeq \epsilon  \int_0^t dt' V[{\bf r}(t'),\bp (t')]$ \cite{JP01}.
 The quantity $w$ is $ \hbar / \sigma$ for sufficiently small $\sigma$ and depends on both $\sigma$ and
 the local instability of the classical trajectory when $\sigma$ is not very small \cite{WL05}.




 We first discuss the SP in the case of $d=1$ with a regular dynamics.
 We assume that the GS can be (approximately) written as a Gaussian wave packet
 in certain coordinate of the classical counterpart.
 This is possible, e.g., in the models discussed below.
 In this case, as shown in Ref.~\cite{wwg-LEr}, for $t>T$,
 due to the periodicity of the classical motion,
 the main contribution of $\Delta S$ to
 the SP is given by its average part $\epsilon U t$, where $U= \frac 1T \int_0^T V(t)dt$
 and $T$ is the period of the classical motion in $H(\lambda)$.
 Upto the first order expansion of $U$ in $p_0$,  Eq.~(\ref{mt-sc}) predicts a
 Gaussian decay of the SP \cite{PZ02,wwg-LEr}.
 For relatively long times, higher order terms of $U$ induces
 power law decay of the SP \cite{wwg-LEr,note-power}.
 For example, to the second order term,
 \be M_{1}(t) \simeq {c_0}{(1+\xi^2 t^2)^{-1/2}}
 e^{ -\Gamma t^2 /(1+\xi^2 t^2)} , \label{Mt-sc} \ee
 where $c_0\sim 1$, $\Gamma = ( \frac{\epsilon  w}{\hbar} \frac{\pp \ww U}{\pp p_0}  )^2/2$,
 $\xi = | \frac{\epsilon w^2}{2\hbar} \frac{ \pp^2 \ww U} { \pp p_0^2}|$,
 with tilde indicating evaluation at $\ww { p}_0$ \cite{wwg-LEr}.
 It is seen that $M_1$ has a Gaussian decay $e^{-\Gamma t^2}$ for initial times and
 has a $1/{\xi t}$ decay for long times.

 Next, we consider the case of a regular classical counterpart with large $d$.
 In this case, the underlying classical motion is typically quasi-periodic with many different
 frequencies, as a result, $T$ is usually much longer than  time scales
 of practical interest.
 For times $t \ll T$, classical trajectories may look random in the torus,
 due to the difference in the frequencies.
 To calculate the SP in this case,
 one may write it in terms of the distribution $P(\Delta S)$ of $\Delta S$
 (with the Gaussian weight taken into account),
 $ M_{\rm sc}(t) \simeq \left | \int d\Delta S e^{i\Delta S/ \hbar }
 P(\Delta S)\right |^2 \label{Mp-ps} $.
 When the trajectories can be effectively regarded as random walks for times $t\ll T$
 due to the many frequencies,
 $P(\Delta S)$ is close to a Gaussian distribution, independent of the initial state.
 In this case, the SP can be calculated in the same way as in a chaotic system \cite{CT02},
 which has an exponential decay determined by  the variance of $\Delta S$,
 \bey   M_2(t)  \simeq e^{-K_s \epsilon^2  t/\hbar^2},  \label{Mt-fgr} \eey
 where
 \be K_s \simeq \frac 1t \left \la \left [ \int Vdt \right ]^2 - \left \la \int Vdt \right
 \ra^2 \right \ra , \ee
 with $\int Vdt =\int_0^t dt' V[{\bf r}(t'),\bp (t')]$ \cite{note-expon}.


 To summarize, for small $\epsilon$ and sufficiently small $\delta$, and for relatively long times,
 the SP may have a power law decay when $d=1$,
 and has the exponential decay $M_2(t)$ when $d$ is sufficiently large.
 We remark that, for $\lambda$ far from $\lambda_c$, the SP
 is always close to 1 for small $\epsilon$.


\begin{figure}
  \vspace{-0.5cm}
 \includegraphics[width=8.0cm]{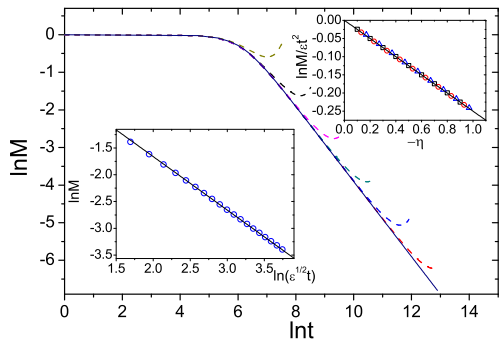}
  \vspace{-0.5cm}
 \caption{(Color online)
 Decay of the SP (dashed curves) in the normal phase of Dicke model.
 Parameters: $\omega =\omega_0=1$, $\epsilon =10^{-5}$, and $\delta =-10^{-m}$
 with $m=6,7,8,9,10,11$ from top to bottom.
 The solid curve is a fitting curve of the form in Eq.~(\ref{Mt-sc}),
 having an initial Gaussian decay $e^{-\Gamma t^2}$ followed by a $1/\xi t$ decay.
 The $1/t$ decay becomes clear with increasing $m$,
 i.e., with $\lambda'$ approaching $\lambda_c$.
 Upper right inset: $(\ln M)/\epsilon t^2$ versus $\eta$ for
 different pairs of $(\epsilon,t)$ with short $t$,
 in agreement with the prediction $\Gamma \sim |\eta\epsilon|$.
 Lower left inset: $\ln M$ versus $\ln(\epsilon^{1/2}t) $ for  $\epsilon \in (10^{-6},10^{-5})$
 and $\ln t\in (8.6,9.5)$ in the $1/t$ decay region.
 $\delta =-10^{-10}$, thus, $|\eta| \simeq 1$.
 The results are in agreement with the prediction $\xi \sim |\eta \epsilon|^{1/2}$.
 }
 \label{fig-dicke}
 \end{figure}

 \section{Numerical simulations}

 The first model we study is the single-mode Dicke model \cite{Dicke54},
 describing the interaction between a single bosonic mode and a collection of $N$ two-level atoms.
 In terms of collective operators ${\bf J}$ for the $N$ atoms,
 the Dicke Hamiltonian is written as (hereafter we take $\hbar =1$) \cite{EB03},
 \be
 H=\omega_{0}J_{z}+\omega a^{\dag}a + ({\lambda}/{\sqrt{N}})(a^{\dag}+a)(J_{+}+J_{-}).
 \label{DH} \ee
 In the limit $N\to \infty$, the system undergoes a QPT at
 $\lambda_c = \frac 12 \sqrt{\omega \omega_0}$, with a normal phase for $\lambda <\lambda_c$
 and a super-radiant phase for $\lambda > \lambda_c$.
 The Hamiltonian can be diagonalized in this limit,
 \be H(\lambda)= \sum_{k=1,2} e_{k\lambda} c_{k\lambda}^{\dag} c_{k\lambda} + g , \ee
 where $c_{k\lambda}^{\dag}$ and $c_{k\lambda}$ are bosonic creation and annihilation
 operators, $e_{k\lambda}$ are single quasi-particle energies,
 and $g$ is a c-number function \cite{EB03}.
 To be specific, in the normal phase,
 \be e_{k\lambda}^{2}=\frac{1}{2}\left \{ \omega^{2}+\omega_{0}^{2} +
 (-1)^k \sqrt{(\omega_{0}^{2}-\omega^{2} )^{2}+16\lambda^{2}\omega\omega_{0}} \right \} .
 \ee
 It is seen that $e_{1\lambda_c}=0$, hence, the ground level of $H(\lambda_c)$ is infinitely degenerate.
 Since $e_{2\lambda_c}= \sqrt{\omega^{2}+\omega_{0}^{2}}$ is finite,
 at the QPT one may consider the effective Hamiltonian $H_{\rm eff}(\lambda) =
 e_{1\lambda}c_{1\lambda}^{\dag}c_{1\lambda}$ with $d=1$.
 Direct calculation shows $e_{1\lambda} \simeq A|\Delta \lambda|^{1/2}$, with
 $A=\frac{2(\omega \omega_0)^{3/4}} {\sqrt{\omega^2+\omega_0^2}}$,
 and
 \be V = -\frac{A^2}{2e_{1\lambda}} \left (c_{1\lambda}^{\dag}c_{1\lambda}
 + 2(c_{1\lambda}^{\dag})^2 + 2c_{1\lambda}^2 \right ) \sim |\Delta \lambda|^{-1/2}.
 \ee

 The semiclassical result Eq.~(\ref{Mt-sc}) predicts that the SP has a Gaussian decay followed by
 a power law decay, with scaling properties
 \be \Gamma \sim \frac{\epsilon^2}{|\Delta \lambda |^{-1}} = |\eta \epsilon |, \hspace{1cm}
 \xi \sim |\eta \epsilon |^{1/2}.
 \ee
 These predictions have been confirmed in our numerical simulations (Fig.~\ref{fig-dicke}).
 Numerically, the SP was calculated by making use of relations between
 $(c_{k\lambda}^{\dag}, c_{k\lambda})$ and $(c_{k\lambda'}^{\dag}, c_{k\lambda'})$,
 which can be directly derived from formulas given in Ref.~\cite{EB03}.
 Our numerical results support the prediction that the semiclassical theory may work for sufficiently
 small $|\delta |$.
 Similar results have also been found in the super-radiant phase.



 The second model we have studied is the LMG model \cite{lipkin}, with the Hamiltonian
 $H=-\frac{1}{N}(S_x^2+{\gamma}S_y^2)-\lambda S_z$, which has a critical point
 at $\lambda_c =1$\cite{Dusuel}.
 The model has a classical counterpart with $d=1$.
 Direct computation shows a $1/t$ decay of the SP for relatively long times
 in the neighborhood of $\lambda_c$ \cite{WZW09}.

 As a third model, we study a 1-dimensional Ising chain in a transverse field,
 \be
 H(\lambda) = -\sum_{i=1}^N \sigma_i^z \sigma_{i+1}^z + \lambda \sigma_i^x .
 \ee
 The Hamiltonian can be diagonalized by using Jordan-Wigner and
 Bogoliubov transformations,
 giving $H(\lambda)=\sum_k e_k (b_k^{\dag}b_k-1/2)$ \cite{Sach99}.
 Here, $b_k^{\dag}$ and $b_k$ are creation and annihilation operators for fermions and
 $e_k$ are single quasi-particle energies,
 \be e_k = 2\sqrt{1+\lambda^2-2\lambda \cos (ka)} \label{ek} \ee
 with lattice spacing $a$,
 where $k=\frac{2\pi m}{aN}$ with $m=-M, -M+1,\ldots , M$ and $N=2M+1$.
 Note that $(ka)$ in Eq.~(\ref{ek}) is in fact independent of the lattice spacing $a$,
 with $ka=2\pi m/N$.

 To understand the degeneracy property of the ground level in the large $N$ limit,
 let us consider those $m$
 satisfying $|m| < N^{\beta}$ for large $N$, where $\beta \in (0,1)$ is an arbitrary number
 independent of $N$.
 In the limit $N\to \infty$, one has $ka \to 0$ for these $m$.
 As a result, Eq.~(\ref{ek}) gives $e_k = 2|\Delta \lambda |$ with $\lambda_c=1$,
 in particular, at the critical point $\lambda =\lambda_c $, $e_k=0$ for these modes $m$.
 The number of these modes $m$ is infinitely large in the limit $N\to \infty$,
 hence, the ground level is infinitely degenerate.

\begin{figure}
 \includegraphics[width=8.0cm]{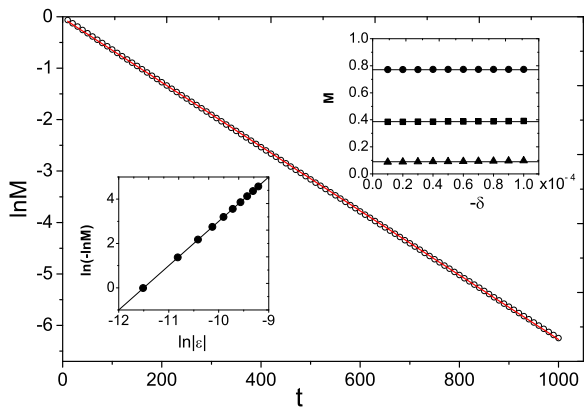}
 \vspace{0.1cm}
 \caption{(Color online)
 Decay of the SP (circles) in a 1-dimensional Ising chain in a transverse field, with $N=2\times 10^8$,
 $\epsilon =8\times 10^{-6}$, and $\delta =-4\times 10^{-6}$.
 It has the expected exponential decay (solid straight line).
 Lower left inset: Dependence of $\ln (-\ln M)$ on $\ln|\epsilon |$ for a fixed time $t$.
 The straight line has a slope 2, as predicted in Eq.~(\ref{Mt-fgr}).
 Upper right inset:
 The SP increases slowly with increasing $|\delta |$ for fixed $\epsilon $ and $t$,
 in agreement with the prediction for $K_s$ given in the text.
 }
 \label{fig-is}
 \end{figure}

 In a sufficiently low energy region and for $\lambda \simeq \lambda_c$,
 a classical counterpart of the system can be introduced as follows.
 For $\lambda =\lambda_c$,
 $e_k \simeq 4\pi |m|/N$ for sufficiently large $N$ and small $|m|$.
 In the low energy region,
 due to this linear dependence of $e_k$ on $m$,
 using the method of bosonization (see Ref.~\cite{Sach99}), one can
 express fermionic states
 $ b_{k_1}^{\dag} \ldots b_{k_n}^{\dag} |{\rm vacuum}\ra $
 in terms of (many) bosonic modes.
 Each bosonic mode has a classical counterpart with one degree of freedom,
 hence, $H(\lambda_c)$ has a classical counterpart in the low energy region with a
 large value of $d$ ($d \to \infty$ in the large $N$ limit).
 This implies that $H(\lambda)$ with $\lambda \simeq \lambda_c$
 also has a classical counterpart with large $d$,
 as a result, typically the SP should have an exponential decay $M_2(t)$ in Eq.~(\ref{Mt-fgr}).

 Direct derivation shows that the perturbation in this model is
 \be V= \frac{\lambda -\cos ka}{e_k/4}(b_k^{\dag}b_k-\frac 12) +
 \frac{\sin ka}{e_k/2} i(b_kb_{-k}
 -b^{\dag}_kb^{\dag}_{-k}) \label{V-is} .
 \ee
 Further analysis shows that $V$ has no singularity at the critical point, e.g.,
 \bey \frac{\sin ka}{e_k} \sim \left \{
 \begin{array}{l}
 \sin ka /|\Delta \lambda| , \ \hspace{1.5cm} \text{for} \ |ka| \lesssim |\Delta \lambda|
 \\ \sin ka /\sqrt{1-\cos ka} , \hspace{0.5cm} \text{for} \ |ka| > |\Delta \lambda|
 \end{array} \right . . \ \label{sinka} \eey
 Therefore, $K_s$ in Eq.~(\ref{Mt-fgr}) has no singularity in the vicinity of $\lambda_c$.
 For large and fixed $N$ and for $|\Delta \lambda |\gg 1/N$,
 since the coupling strength of $V$ in the eigenbasis of $H(\lambda)$
 increases with decreasing $|\Delta \lambda|$,
 $K_s$ should increase slowly with decreasing $|\Delta \lambda|$.


 Numerical computation of the SP can be done by using the following expression
 given in Ref.\cite{Quan06},
 \be M(t) = \prod_{k>0} F_k,  \ee
 where
 \bey F_k = 1-\sin^2(\theta_\lambda-\theta_{\lambda'}) \sin^2(e_kt),
 \\ \theta_\lambda = \arctan\frac{-\sin (ka)}{\cos (ka) -\lambda},
 \eey
 and $e_k$ are evaluated at $\lambda'$ .
 Our numerical computations confirm not only the prediction of an exponential decay
 of the SP at the criticality, but also some details in the exponent of
 $M_2(t)$ discussed above (see Fig.\ref{fig-is}),
 namely, the $\epsilon^2$ dependence and the properties of $K_s$.

 As a fourth model, we have studied the XY model \cite{Sach99}, with the Hamiltonian
 \be H=-\sum_{i}\frac{1+\gamma}{2}\sigma_i^x\sigma_{i+1}^x+
 \frac{1-\gamma}{2}\sigma_i^y\sigma_{i+1}^y+\frac{\lambda}{2}\sigma_i^z,
 \ee
 which has critical points $\lambda_c = \pm 1$.
 As in the Ising chain, in the low energy region around $\lambda_c$,
 the XY model has a classical counterpart with large $d$.
 The SP in this model can be calculated in a way similar to that in the Ising chain discussed above,
 and our numerical simulations also confirmed the semiclassically predicted
 exponential decay of the SP.

 \section{Conclusions and discussions}

 We have shown that the semiclassical theory may be used in the study of
 the decay of SP (survival probability) of GS (ground states)
 in the vicinity of those QPT with infinitely degenerate ground levels at the critical points.
 Two qualitatively different decaying behaviors of the SP
 have been found for relatively long times: power law decay in systems with $d=1$
 and exponential decay in systems with sufficiently large $d$, where $d$ is the
 degrees of freedom of the classical counterpart of the quantum system.

 The above results suggest that the SP decay may be useful in the classification of QPT,
 an important topic far from being completely solved, in particular, in the non-equilibrium regime.
 Here, we have found two classes: one class with power law decay and another class with
 exponential decay.
 It needs further investigation whether other types of SP decay may appear
 at QPT, e.g., relatively-long-time Gaussian decay or a decay between power-law
 and exponential.

 W.-G.W. is grateful to P.Braun, F.Haake, R.Sch\"{u}tzhold, J.Gong,
 T.Prosen, G.Benenti, and G.Casati for helpful discussions.
 This work is partly supported by the Natural Science Foundation of China under Grant Nos.~10775123
 and 10975123 and the National Fundamental Research Programme of China
 Grant No.2007CB925200.



\begin{thebibliography}{99}
\bibitem{foot-osc} In systems with finite $N$, the SP may show a periodic behavior
 for sufficiently long times \cite{Quan06}.
 Here, we study the decay law of the SP for relatively long times in the thermodynamic limit.
\bibitem{note-power}
 The Gaussian shape of the initial state is irrelevant in the derivation of the power law
 decay of $M_L(t)$.
 Moreover, as shown in \cite{PZ02}, $M_L(t)$ has a long-time $t^{-d}$ decay
 for initial random states.
 These imply that the power law feature might be insensitive to the initial condition.
\bibitem{note-expon} Dividing $\ov M_L(t)$ in regular systems
 into diagonal and off-diagonal parts, Ref.~\cite{JAB03} shows that the off-diagonal part
 may have an exponential decay under certain condition.
\end{thebibliography}
 \end{document}